\documentclass[reprint,
 amsmath,amssymb,
 aps,
pre,
]{revtex4-1}

\usepackage{xcolor}
\usepackage{physics}
\usepackage{graphicx}
\usepackage{dcolumn}
\usepackage{bm}
\usepackage{hyperref}
\usepackage{verbatim}

\newcommand{\Ut}{{\widetilde{U}}}
\newcommand{\Et}{{\widetilde{E}}}
\newcommand{\Wt}{{\widetilde{W}}}
\newcommand{\Zt}{{\widetilde{Z}}}
\newcommand{\Tt}{{\widetilde{T}}}
\newcommand{\gt}{{\widetilde{g}}}
\newcommand{\Ec}{{\mathcal{E}}}

\newcommand{\bq}{{\bf q}}
\newcommand{\bp}{{\bf p}}

\newcommand{\br}{{\bf r}}

\begin{document}


\title{Efficient implementation of the Wang-Landau algorithm for systems with length-scalable potential energy functions}

\author{Santosh Kumar}
\email{skumar.physics@gmail.com}
\author{Girish Kumar}
 \author{Rohit S. Chandramouli}
 \author{Shashank Anand}
 \affiliation{Department of Physics, Shiv Nadar University, Gautam Buddha Nagar, Uttar Pradesh-201314, India }%


\begin{abstract}
We consider a class of systems where $N$ identical particles with positions $\bq_1,...,\bq_N$ and momenta $\bp_1,...,\bp_N$ are enclosed in a box of size $L$, and exhibit the scaling $\mathcal{U}(L\br_1,...,L\br_N)=\alpha(L)\, \mathcal{U}(\br_1,...,\br_N)$ for the associated potential energy function $\mathcal{U}(\bq_1,...,\bq_N)$. For these systems, we propose an efficient implementation of the Wang-Landau algorithm for evaluating thermodynamic observables involving energy and volume fluctuations in the microcanonical description, and temperature and volume fluctuations in the canonical description. This requires performing the Wang-Landau simulation in a scaled box of unit size and evaluating the density of states corresponding to the potential energy part only. To demonstrate the efficacy of our approach, as example systems, we consider Padmanabhan's binary star model and an ideal gas trapped in a harmonic potential within the box. 
\end{abstract}

\pacs{Valid PACS appear here}
\maketitle

\section{Introduction}

The Wang-Landau algorithm is a powerful Monte Carlo technique that gives direct access to the density of states of statistical models of interest~\cite{WL2001a,WL2001b}. It is based on performing a random walk in the energy space and obtaining the density of states, up to a multiplicative constant, in an iterative manner. The Wang-Landau algorithm was originally used to study systems exhibiting discrete energy levels. Owing to several refinements and improvements~\cite{YFP2002,SDP2002,SBML2003,Calvo2002,CP2003,ZB2005,JSM2005,AE2006,ZSTL2006,LOL2006,PCABD2006,GC2007,GHH2007,BP2007,BMP2008,CCTDL2008,JBJ2008,GWLX2009,WLL2011,JP2016,DD2012,VLWL2014,KBSM2016,SZJ2016,SZBIP2017,CBR2017}, it has been gradually implemented to study complicated systems with continuous energy spectra as well. Examples include complex fluids~\cite{YFP2002,SDP2002,GC2007}, atomic clusters~\cite{Calvo2002,CP2003}, liquid crystals~\cite{JSM2005}, biomolecules~\cite{PCABD2006,GHH2007,GWLX2009}, polymers~\cite{CCTDL2008,WLL2011,JBJ2008,JP2016}, logarithmic gas in the context of random matrix theory~\cite{SK2013}, etc.

Usually one obtains the density of states as a function of energy alone by keeping the number of particles and the volume of the system fixed. Consequently, the thermodynamic observables that depend on particle number and volume fluctuations cannot be evaluated. In Refs.~\cite{YFP2002,SDP2002} the authors proposed an off-lattice generalization of the original scheme that enables one to obtain the density of states as a function of energy, number of particles, and volume. This requires performing the Wang-Landau simulation in a three-dimensional space. Subsequently, the extensions to this have also been considered, see, for example, Refs.~\cite{AE2006, DD2012}. These generalizations, however, come at the expense of increased computational power and time. 

In the present paper, we consider a system comprising a fixed number of particles enclosed inside a box of arbitrary volume. Within the box, the system exhibits a certain scaling behavior for its potential energy function. We show how the Wang-Landau algorithm can be used for this system to calculate thermodynamic observables involving energy and volume fluctuations in the microcanonical ensemble and temperature and volume fluctuations in the canonical ensemble. Interestingly, this involves performing the Wang-Landau simulation in a one-dimensional space only and obtaining the density of states for potential energy, for the case of a scaled box of unit size. To demonstrate the efficacy of our approach, as example systems, we consider Padmanabhan's binary star model~\cite{Padmanabhan1990} and a noninteracting gas trapped in a harmonic potential within a container. The former exhibits peculiar behavior, such as inequivalence of the statistical ensembles, due to long-range interactions. The latter, being a system of noninteracting particles, does not exhibit any peculiarities, and the microcanonical and canonical descriptions agree.
 
The presentation scheme in this paper is as follows. In Sec.~\ref{SecStatEns}, we collect some standard results concerning the microcanonical and the canonical ensembles and the corresponding thermodynamic observables. In Sec.~\ref{SecScaling}, we describe the scaling behavior exhibited by the potential energy function, which is essential for our approach to work. Moreover, based on this scaling, we reexpress the results laid out in the preceding section to forms suited for implementation of the Wang-Landau simulation. Section~\ref{SecExamps} deals with the two models described above for validating our results. Finally, we conclude in Sec.~\ref{SecSum} with a brief summary of our proposed scheme and results.

\section{Statistical ensembles and Thermodynamic observables}
\label{SecStatEns}

We consider a system of $N$ identical particles of mass $m$ with positions ${\bf q}_j$ and momenta ${\bf p}_j$, where $j=1,...,N$. The Hamiltonian for this system is given by
\begin{equation}
\label{Hqp}
\mathcal{H}(\bq_1,...,\bq_N; \bp_1,...,\bp_N)=\sum_{j=1}^N \frac{\bp_j^2}{2m}+\mathcal{U}(\bq_1,...,\bq_N).
\end{equation}
Here the first term on the right hand side is the kinetic energy part, and the second term constitutes the potential energy.
Additionally, we constrain the position coordinates of the particles by enclosing the system in a cubical box of edge length $L$ or a spherical box of radius $L$. Effectively, this amounts to introducing an infinite potential region outside the box boundary and therefore adding another potential energy term in the Hamiltonian above, which is zero inside the box and infinity outside. For simplicity, we do not explicitly show this term in the Hamiltonian. 

The calculation of thermodynamic observables associated with the system described by the Hamiltonian $\mathcal{H}$ requires obtaining certain key quantities, such as entropy for the microcanonical ensemble and partition function for the canonical ensemble~\cite{Huang1987,PB2011}. In the evaluation of either of these, it is possible to perform the integral over the momentum-space variables analytically. Consequently, the analysis boils down to the calculation of the density of states $\mathcal{G}(U,V)$ for the potential energy $U$,
\begin{equation}
\label{Guvn}
\mathcal{G}(U,V)=\int \cdots \int \prod_{j=1}^N d^3 q_j\, \delta [U-\mathcal{U}({\bf q}_1,...,{\bf q}_N)].
\end{equation}
Here $\delta(\cdot)$ represents the Dirac $\delta$ function, and the integration is over the volume $V$ of the box. Since we are concerned with fixed $N$, for simplicity of notation, we have suppressed its appearance as an argument in $\mathcal{G}$. We do the same for other functions appearing below. 

The Wang-Landau algorithm can be used to calculate $\mathcal{G}(U,V)$, albeit up to a multiplicative constant. This constant is, however, inconsequential in calculating thermodynamic quantities. In practice, numerical simulations aim to obtain the logarithm of the density of states (up to an additive constant) to prevent overflow in computation.
 
\subsection{Microcanonical ensemble}

In the microcanonical ensemble approach, the key quantity is entropy of the system~\cite{Huang1987,PB2011}, given by
\begin{align}
S(E,V)=k_B \ln\Omega(E,V).
\end{align}
Here, $k_B$ is the Boltzmann constant, $h$ is the Planck's constant, and $\Omega(E,V)$ is the density of states, viz.,
\begin{align}
\label{Gevn}
\Omega(E,V)=\frac{1}{N!\, h^{3N}}\int \cdots \int \prod_{j=1}^N d^3 p_j d^3 q_j\, \delta(E-\mathcal{H}).
\end{align}
It should be noted that, unlike $\mathcal{G}(U,V)$ of Eq.~\eqref{Guvn}, a direct evaluation of $\Omega(E,V)$ itself from algorithms such as Wang-Landau is impractical. This is because the random walk in $E$ space would require evaluations using the points from the full phase space (positions as well as momenta). Moreover, the momentum coordinates are unbounded. Therefore, typically, such simulation schemes are used for ``conformational" microcanonical ensembles which are defined solely using the potential energy. For instance, classical spin systems where the kinetic energy is usually not defined in the Hamiltonian. However, here we are concerned with the ``real" microcanonical ensemble wherein the kinetic energy part is also considered~\cite{SZJ2016,SZBIP2017}.

The integration over the momentum variables can be performed analytically in~\eqref{Gevn} and leads to 
\begin{align}
\label{S1}
S(E,V)=k_B \ln(\frac{(2\pi m)^{3N/2}}{N! \,\Gamma(3N/2)h^{3N}})+k_B\ln W(E,V),
\end{align}
where, with $\Theta(\cdot)$ being the Heaviside $\Theta$ function, 
\begin{align}
\label{MCanSW}
\nonumber
&W(E,V)=\int\cdots \int \prod_{j=1}^N d^3 q_j (E-\mathcal{U})^{3N/2-1} \Theta\big(E-\mathcal{U}),\\
&~~~~=\int d U \mathcal{G}(U,V)\, (E-U)^{3N/2-1}\,\Theta(E-U).
\end{align}
Within the microcanonical approach, the desired thermodynamic quantities then follow through the standard relations. For instance, temperature ($T$) and pressure ($P$) are obtained using
\begin{equation}
\label{MCan}
\frac{1}{T}=\left(\frac{\partial S}{\partial E}\right)_V,~~~\frac{P}{T}=\left(\frac{\partial S}{\partial V}\right)_E.
\end{equation}
As discussed in the Introduction, in general, to calculate quantities that involve number and volume fluctuations along with energy fluctuations, we need to obtain $\mathcal{G}$ in its {\it full} generality as a function of $U$, $V$, and $N$. In principle, the Wang-Landau approach can be used to numerically obtain the general density of states by performing a simulation in a three-dimensional space~\cite{YFP2002,SDP2002}. However, its implementation is often not possible for complex systems, and even if possible, it is computationally very demanding. Typically, one performs the simulation to explore the $U$ space only by fixing $V$ and $N$, and obtains the density of states $g(U)$ for the potential energy. In such a situation, one can calculate the quantities which involve energy fluctuations only, i.e., the ones that follow from the partial derivative with respect to $E$, e.g., the temperature $T$ in Eq.~\eqref{MCan}. 

\subsection{Canonical ensemble}

For the canonical ensemble, the central object is the partition function~\cite{Huang1987,PB2011},
\begin{equation}
Z(T,V)=\frac{1}{N!\, h^{3N}}\int\cdots \int \prod_{j=1}^N d^3p_j d^3q_j \exp(-\frac{\mathcal{H}}{k_BT}),
\end{equation}
which can be interpreted as the Laplace transform of $\Omega(E,V)$ from $E$ to $1/(k_BT)$ space~\cite{PB2011}.
The momenta integrals can again be performed using standard techniques, leading to
\begin{align}
\label{CanSW}
\nonumber
Z(T,V)&=\frac{\left(2\pi m k_B T\right)^{3N/2} }{N!\, h^{3N}}\int\cdots \int \prod_{j=1}^N d^3 q_j \exp(-\frac{\mathcal{U}}{k_B T})\\
\nonumber
&=\frac{\left(2\pi m k_B T\right)^{3N/2} }{N!\, h^{3N}}\int dU \mathcal{G}(U,V)\exp(-\frac{U}{k_BT})\\
&\equiv (2\pi m k_B T)^{3N/2} Z_U(T,V).
\end{align}
We defined here $Z_U$ as the partition function corresponding to the potential energy. The Helmholtz free energy, given by
\begin{equation}
F(T,V)=-k_B T \ln Z(T,V)
\end{equation}
leads to other thermodynamic quantities of interest. For instance, the pressure and entropy are given, respectively, by
\begin{equation}
\label{Can}
P=-\left(\frac{\partial F}{\partial V}\right)_T,~~~~S=-\left(\frac{\partial F}{\partial T}\right)_V.
\end{equation}
The average energy is obtained using
\begin{align}
\label{avE}
E&=k_BT^2\frac{\partial \ln Z}{\partial T}
=\frac{3}{2}Nk_B T+k_BT^2\frac{\partial \ln Z_U}{\partial T},
\end{align}
where the first term is the kinetic-energy contribution and the second term gives the potential energy contribution. In comparing the results from the microcanonical ensemble with those of the canonical ensemble, we identify  the constant energy at which the former is defined with the mean energy corresponding to the latter. Therefore, for simplicity of notation we use $E$ for the average energy in the canonical ensemble instead of a more appropriate notation of $\expval{E}$.

In a similar fashion to the microcanonical approach, to obtain quantities that involve temperature, volume, and number fluctuations, we need $\mathcal{G}$ as function of $U,V$, and $N$. However, in this case, calculation of quantities involving partial derivatives with respect to $T$ is straightforward, and for these one needs to sample only the $U$ space and obtain $g(U)$.

For the Hamiltonian given by Eq.~\eqref{Hqp}, we suggest below a methodology which enables us to calculate quantities which involve partial derivative with respect to volume, in addition to energy or temperature for the microcanonical and canonical ensembles, respectively. This requires performing the random walk in potential energy space only. However, as we see below, for this approach to work, the potential energy function needs to exhibit a certain scaling behavior.

\section{Scaling behavior for the Potential Energy function}
\label{SecScaling}

We now focus on a class of systems where the potential energy, within the box of size $L$, exhibits the scaling property,
\begin{align}
\label{scaling}
\mathcal{U}(L \,\br_1,...,L\, \br_N)=\alpha\, \mathcal{U}(\br_1,...,\br_N),
\end{align}
where $\alpha\equiv \alpha(L)$ and $\br_j$ are the scaled positions. Equivalently, this scaling behavior can also be expressed in terms of the volume since $V\sim L^3$.
Examples include noninteracting particles in a power-law trap,
\begin{align}
\label{trap}
\mathcal{U}({\bf q}_1,...,{\bf q}_N)\sim \sum_j |\bq_j|^\gamma,
\end{align}
for which $\alpha=L^\gamma\sim V^{\gamma/3}$; particles with van der Waals interaction involving a hard core,
\begin{align}
\label{VWall}
\mathcal{U}({\bf q}_1,...,{\bf q}_N)\sim\begin{cases}
  -\displaystyle\sum_{j\neq k}\dfrac{1}{|\bq_j-\bq_k|^6}, & |\bq_j-\bq_k|>\epsilon L,\\
  ~~~+\infty, & |\bq_j-\bq_k|< \epsilon L,
\end{cases}
\end{align}
where $\epsilon$ is a small dimensionless constant, giving $\alpha=1/L^6\sim 1/V^2$; and the Newtonian-Coulombic potential energy
\begin{equation}
\label{NC}
\mathcal{U}({\bf q}_1,...,{\bf q}_N)\sim -\sum_{j\neq k}\frac{1}{|{\bf q}_j-{\bf q}_k|},
\end{equation}
for which $\alpha=1/L\sim1/V^{1/3}$. A small distance cut off can be introduced in this case also to prevent {\it collapse}~\cite{Padmanabhan1990}.

Given the scaling property~\eqref{scaling}, it turns out that we need to obtain only the density of states $\gt(\Ut)$ for the potential energy $\Ut=U/\alpha$ with the system now confined to size 1, i.e., a cubical box of edge-length 1 and hence volume 1, or a spherical container of radius 1 and therefore volume $4\pi/3$. We have
\begin{equation}
\label{gtld}
\gt(\Ut)=\int \cdots \int \prod_{j=1}^N d^3 r_j\, \delta \Big(\Ut-\mathcal{U}(\br_1,...,\br_N)\Big).
\end{equation}
Although the simulation is performed with unit volume (fixed), calculating the above would give access to quantities that depend on volume fluctuations, i.e., involve derivatives with respect to volume.  We see this below separately for the microcanonical and canonical ensembles.
 
\subsection{Microcanonical ensemble}
In this case, Eq.~\eqref{MCanSW} leads to
\begin{align}
W(E,V)=L^{3N}\alpha^{3N/2-1} \,\Wt\left(\frac{E}{\alpha}\right),
\end{align}
with $V=L^3$ for the cubical box and $V=4\pi L^3/3$ for the spherical box. The scaling behavior of $W$ may be written in terms of $V$ using the replacement $L\rightarrow(V/c)^{1/3}$. We defined here, $c=1$ for the cubical box and $c=4\pi/3$ for the spherical box. The function $\Wt$ in the above equation is given by
\begin{align}
\Wt(\Et)=\int d\Ut \gt(\Ut)\big(\Et-\Ut)^{3N/2-1} \Theta\big(\Et-\Ut \big),
\end{align}
with
$\Et=E/\alpha$ being the scaled total energy. It should be noted that if $\mathcal{U}$ contains some constants, they may also be included in the definition of $\Et$ thereby making those constants in the expression of $\mathcal{U}$ effectively unity. The temperature and pressure can now be calculated as
\begin{align}
\label{Tinv}
\frac{1}{T}=\frac{k_B}{\alpha}\,\phi\left(\frac{E}{\alpha}\right),
\end{align}
and
\begin{align}
\label{PbyT}
\nonumber
\frac{P}{T}&=\frac{k_B}{3cL^2}\Bigg[\frac{3N}{L}+\left(\frac{3N}{2}-1\right)\frac{1}{\alpha}\frac{\partial\alpha}{\partial L}
-\frac{E}{\alpha^2}\frac{\partial\alpha}{\partial L}\,\phi\left(\frac{E}{\alpha}\right)\Bigg]\\
&=\frac{k_B}{3cL^2}\Bigg[\frac{3N}{L}+\left(\frac{3N}{2}-1\right)\frac{1}{\alpha}\frac{\partial\alpha}{\partial L}\Bigg]
-\frac{E}{3cL^2\alpha}\frac{\partial\alpha}{\partial L}\frac{1}{T}.
\end{align}
We have defined here
\begin{align}
\label{phi}
\nonumber
&\phi(\Et)=\frac{\partial \ln \Wt}{\partial \Et}=\frac{1}{\Wt}\frac{\partial \Wt}{\partial \Et}\\
&=\Big(\frac{3N}{2}-1\Big)\frac{\displaystyle\int d\Ut \gt(\Ut)\, (\Et-\Ut)^{3N/2-2}\,\Theta(\Et-\Ut)}{\displaystyle\int d\Ut \gt(\Ut)\, (\Et-\Ut)^{3N/2-1}\,\Theta(\Et-\Ut)}.
\end{align}
We note that any multiplicative constant appearing with $\gt(\Ut)$ cancels from the numerator and the denominator and hence does not alter the value of $\phi(\Et)$. Numerical simulations produce $\gt(\Ut)$ for discretized potential energy space, therefore we need to replace integrals by summations in the above equation and use instead,
\begin{equation}
\label{phisum}
\phi(\Et)=\Big(\frac{3N}{2}-1\Big)\frac{\sum_{j}\gt(\Ut_j)(\Et-\Ut_j)^{3N/2-2}\,\Theta(\Et-\Ut_j)}{\sum_{j}\gt(\Ut_j)(\Et-\Ut_j)^{3N/2-1}\,\Theta(\Et-\Ut_j)}.
\end{equation}
The common factor of $\Delta \Ut$ has been canceled from the sums in the numerator and denominator in the above equation, assuming that the investigated potential energy window is divided into bins of equal size.

\subsection{Canonical ensemble}

We obtain, in this case,
\begin{align}
Z(T,V)=L^{3N}\alpha^{3N/2} \,\Zt\left(\frac{T}{\alpha}\right),
\end{align}
where
\begin{align}
\Zt(\Tt)=\frac{(2\pi mk_B \Tt)^{3N/2}}{N!\, h^{3N}}\int d\Ut \gt(\Ut)\exp(-\frac{\Ut}{k_B \Tt}),
\end{align}
with
$\Tt=T/\alpha$ being the scaled temperature. In this case also, the scaling property can be expressed in terms of the volume. The pressure and entropy follow as
\begin{align}
\label{Pt}
P=\frac{k_BT}{3cL^2}\Big[\frac{3N}{L}+\frac{3N}{2}\frac{1}{\alpha}\frac{\partial\alpha}{\partial L} -\frac{T}{\alpha^2}\frac{\partial\alpha}{\partial L}\,\psi\left(\frac{T}{\alpha}\right)\Big].
\end{align}
\begin{align}
S=k_B \ln\left[L^{3N}\alpha^{3N/2}\Zt\left(\frac{T}{\alpha}\right)\right]+\frac{k_B T}{\alpha}\,\psi \left(\frac{T}{\alpha}\right).
\end{align}
The average energy is
\begin{equation}
\label{Et}
E=\frac{k_BT^2}{\alpha}\,\psi \left(\frac{T}{\alpha}\right).
\end{equation}
The $\psi$ function in the above equations is defined as
\begin{align}
\label{psi}
\nonumber
&\psi(\Tt)=\frac{\partial \ln \Zt}{\partial \Tt}=\frac{1}{\Zt}\frac{\partial \Zt}{\partial \Tt}\\
&=\frac{3N}{2\Tt}+\frac{1}{k_B\Tt^2}\frac{\int d\Ut \gt(\Ut)\,\Ut\exp(-\frac{\Ut}{k_B \Tt})}{\int d\Ut \gt(\Ut)\exp(-\frac{\Ut}{k_B \Tt})}.
\end{align}
Here, the first term corresponds to the kinetic energy contribution, and the second one corresponds to the potential energy. Similar to Eq.~\eqref{phisum}, in practice, we need to use the discretized version of the above equation, namely,
\begin{equation}
\label{psisum}
\psi(\Tt)=\frac{3N}{2\Tt}+\frac{1}{k_B\Tt^2}\frac{\sum_{j}\gt(\Ut_j)\,\Ut_j\,\exp(-\frac{\Ut_j}{k_B \Tt})}{\sum_{j}\gt(\Ut_j)\,\exp(-\frac{\Ut_j}{k_B \Tt})}.
\end{equation}

\section{Example systems}
\label{SecExamps}

In this section, we validate the results laid out in the preceding section for two model systems. As the first system, we consider Padmanabhan's binary star model for which analytical results are available~\cite{Padmanabhan1990}. The second system comprises noninteracting particles in a harmonic trap. For this system, the Wang-Landau results are compared with those from conventional Monte Carlo schemes. In both cases, the system is confined within a box as described in Sec.~\ref{SecStatEns}.

\subsection{Self gravitating binary star}

It is well acknowledged that systems involving long-range interactions exhibit inequivalence of statistical ensembles and peculiar aspects, such as negative specific heat in the microcanonical description~\cite{Padmanabhan1990, Thirring1970, DRAW2002,BB1977,VS2000,VS2002,DRAW2002,Chavanis2002,BB2006,CDR2009}. Padmanabhan has demonstrated that even a toy model, such as the self-gravitating system of two particles, exhibits several peculiarities that are characteristic of self-gravitating systems~\cite{Padmanabhan1990}. Since exact analytical results are available for this model, it serves as a benchmark to test some of the results laid out in the preceding sections.

We consider a system of two bodies ($N=2$), each having mass $m$ and phase space coordinates ${\bq}_1,{\bp}_1$, and ${\bq}_2, {\bp}_2$, respectively. They interact via the long-range gravitational potential. Moreover, the system is enclosed in a spherical box of radius $L$, thereby setting a long distance cutoff. The center of the sphere is chosen to coincide with ${\bq}_1, {\bq}_2=\boldsymbol{0}$. Similarly, a repulsive short distance cutoff is set at a separation $\epsilon L\ll L$, where $\epsilon$ is a small dimensionless parameter. Therefore, the two bodies can be considered as hard spheres of radius $\epsilon L/2$. Consequently, the potential energy function defined within the box is
 \begin{equation}
 \label{UBS}
\mathcal{U}(\bq_1,\bq_2)=\begin{cases}
  -\dfrac{Gm^2}{|\bq_1-\bq_2|}, & |\bq_1-\bq_2|>\epsilon L,\\
  ~~~+\infty, & |\bq_1-\bq_2|<\epsilon L.
 \end{cases}
 \end{equation}
Clearly, in this case, $\alpha=1/L$. Exact analytical results can be derived for this two-body system~\cite{Padmanabhan1990}. In the microcanonical approach, we obtain for Eq.~\eqref{MCanSW},
\begin{align}
\label{BstarW}
 W(E,V)\propto 
\begin{cases}
\displaystyle -\frac{L^4}{\zeta}(1+\epsilon \zeta )^3 ,~~~~~~~~~~~ -\frac{1}{\epsilon}<\zeta<-1,\\
\displaystyle -\frac{L^4}{\zeta}(1+\epsilon \zeta )^3+\frac{L^4}{\zeta}(1+\zeta)^3,~~ \zeta>-1,
\end{cases}
\end{align}
where $\zeta=EL/(Gm^2)$.
Therefore, the temperature and pressure expressions follow using Eq.~\eqref{MCan}: 
\begin{align}
\label{Tmc}
T=\begin{cases}
\displaystyle\frac{E}{k_B}\left(\epsilon \zeta+1\right)\left(2\epsilon\zeta-1\right)^{-1}, ~ -\frac{1}{\epsilon}<\zeta<-1,\\
\displaystyle\frac{Gm^2}{k_B L}[(1-\epsilon^3)\zeta^2+3(1-\epsilon^2)\zeta+3(1-\epsilon)]\\
\times [2(1-\epsilon^3)\zeta+3(1-\epsilon^2)]^{-1},~~~~~ \zeta>-1,
\end{cases}
\end{align}
\begin{align}
P=\begin{cases}
\label{Pmc}
\displaystyle\frac{3E}{4\pi L^3}\left(2\epsilon\zeta+1\right)\left(2\epsilon\zeta-1\right)^{-1},~ -\frac{1}{\epsilon}<\zeta<-1,\\
\displaystyle\frac{3Gm^2}{4\pi L^4}[2(1-\epsilon^3)\zeta^2+5(1-\epsilon^2)\zeta+4(1-\epsilon)]\\
\times [2(1-\epsilon^3)\zeta+3(1-\epsilon^2)]^{-1},~~~~~~~ \zeta> -1.
\end{cases}
\end{align}

For the canonical ensemble, an exact result for the partition function can be written in terms of exponential integral function, $\text{Ei}(u)=-\int_{-u}^\infty dt\, e^{-t}/t$~\cite{GR2007}. We obtain
\begin{align}
\label{BstarZ}
&Z(T,V)\propto \frac{L^3}{\eta^3} \left[ \left(\eta^2+\eta+2\right)\exp(\eta)-\eta^3\text{Ei}(\eta)\right]\\
\nonumber
& -\frac{L^3 \epsilon^3}{\eta^3} \left[\left(\frac{\eta^2}{\epsilon^2}+\frac{\eta}{\epsilon}+2\right)\exp(\frac{\eta}{\epsilon})-\frac{\eta^3}{\epsilon^3}\text{Ei}\left(\frac{\eta}{\epsilon}\right)\right],
\end{align}
where $\eta=Gm^2/(L k_B T)$.
This can be used to calculate exact analytical results for pressure and average energy using Eqs.~\eqref{Can} and~\eqref{avE},
\begin{align}
\nonumber
&P=\frac{3k_B T}{4\pi L^3}\bigg(\left[\left(\eta^2+\eta+4\right)\exp(\eta)-\eta^3\text{Ei}(\eta)\right]\\
\nonumber
& -\epsilon^3\left[\left(\frac{\eta^2}{\epsilon^2}+\frac{\eta}{\epsilon}+4\right)\exp(\frac{\eta}{\epsilon})-\frac{\eta^3}{\epsilon^3}\text{Ei}\left(\frac{\eta}{\epsilon}\right)\right]\bigg)\\
\nonumber
& \times\bigg(\left[\left(\eta^2+\eta+2\right)\exp(\eta)-\eta^3\text{Ei}(\eta)\right]\\
\nonumber
& -\epsilon^3\left[\left(\frac{\eta^2}{\epsilon^2}+\frac{\eta}{\epsilon}+2\right)\exp(\frac{\eta}{\epsilon})-\frac{\eta^3}{\epsilon^3}\text{Ei}\left(\frac{\eta}{\epsilon}\right)\right]\bigg)^{-1},
\end{align}
\begin{align}
\nonumber
&E=6k_B T \left[\exp(\eta)-\epsilon^3 \exp(\frac{\eta}{\epsilon})\right]\\
\nonumber
& \times\bigg(\left[\left(\eta^2+\eta+2\right)\exp(\eta)-\eta^3\text{Ei}(\eta)\right]\\
\nonumber
& -\epsilon^3\left[\left(\frac{\eta^2}{\epsilon^2}+\frac{\eta}{\epsilon}+2\right)\exp(\frac{\eta}{\epsilon})-\frac{\eta^3}{\epsilon^3}\text{Ei}\left(\frac{\eta}{\epsilon}\right)\right]\bigg)^{-1}.
\end{align}

To compare these analytical results with numerics, we use Eqs.~\eqref{Tinv},~\eqref{PbyT} and~\eqref{Pt},~\eqref{Et}, with $N=2$. Within the microcanonical approach, we have
\begin{equation}
\frac{1}{T}=k_B L\,\phi(L E),
\end{equation} 
\begin{equation}
\frac{P}{T}=\frac{k_B}{4\pi L^2}\left(\frac{4}{L}+E\,\phi(L E)\right).
\end{equation}
The equation of state can be obtained from these two relations as
\begin{equation}
PV=\frac{4k_BT}{3}+\frac{E}{3},
\end{equation}
where $V=4\pi L^3/3$. 
With the canonical approach, the pressure is given by
\begin{equation}
P=\frac{k_B T}{4\pi L^2}\left(\frac{3}{L}+T\,\psi(LT)\right),
\end{equation}
and the average energy is obtained as
\begin{equation}
E=k_B T^2 L\,\psi(LT).
\end{equation}
The corresponding equation of state is
\begin{equation}
PV=k_B T+\frac{E}{3}.
\end{equation}

For this system, we considered the spherical box of unit radius ($L=1$), and short distance cut-off $\epsilon L$ with $\epsilon=10^{-3}.$ The $\Ut$-energy window to be explored in the simulation is $[-1000,-0.5]$, where we set $G$, $k_B$, and $m$ equal to unity. The lowest potential energy is decided by the closest possible distance ($=10^{-3}$) between the two particles, whereas the largest is decided by the longest possible distance ($=2$) in the unit sphere. We divided this energy window in 2000 bins and obtained the density of states $\gt(\Ut)$ using the $t$-inverse variant of the Wang-Landau scheme~\cite{BP2007,BMP2008}, with a final modification factor of value $\exp(5\times10^{-7})$. For the simulation, we started by randomly placing the two particles inside the box. Subsequent configurations were generated by randomly perturbing one of the particles by an amount between $-0.01$ and $0.01$, keeping in view the box boundary. The corresponding potential energy values were used in the Wang-Landau implementation. The simulation took about 15 min on a laptop with a 2.8 GHz Intel Core i5 processor. 

\begin{figure}[!t]
\centering
\includegraphics[width=0.7\linewidth]{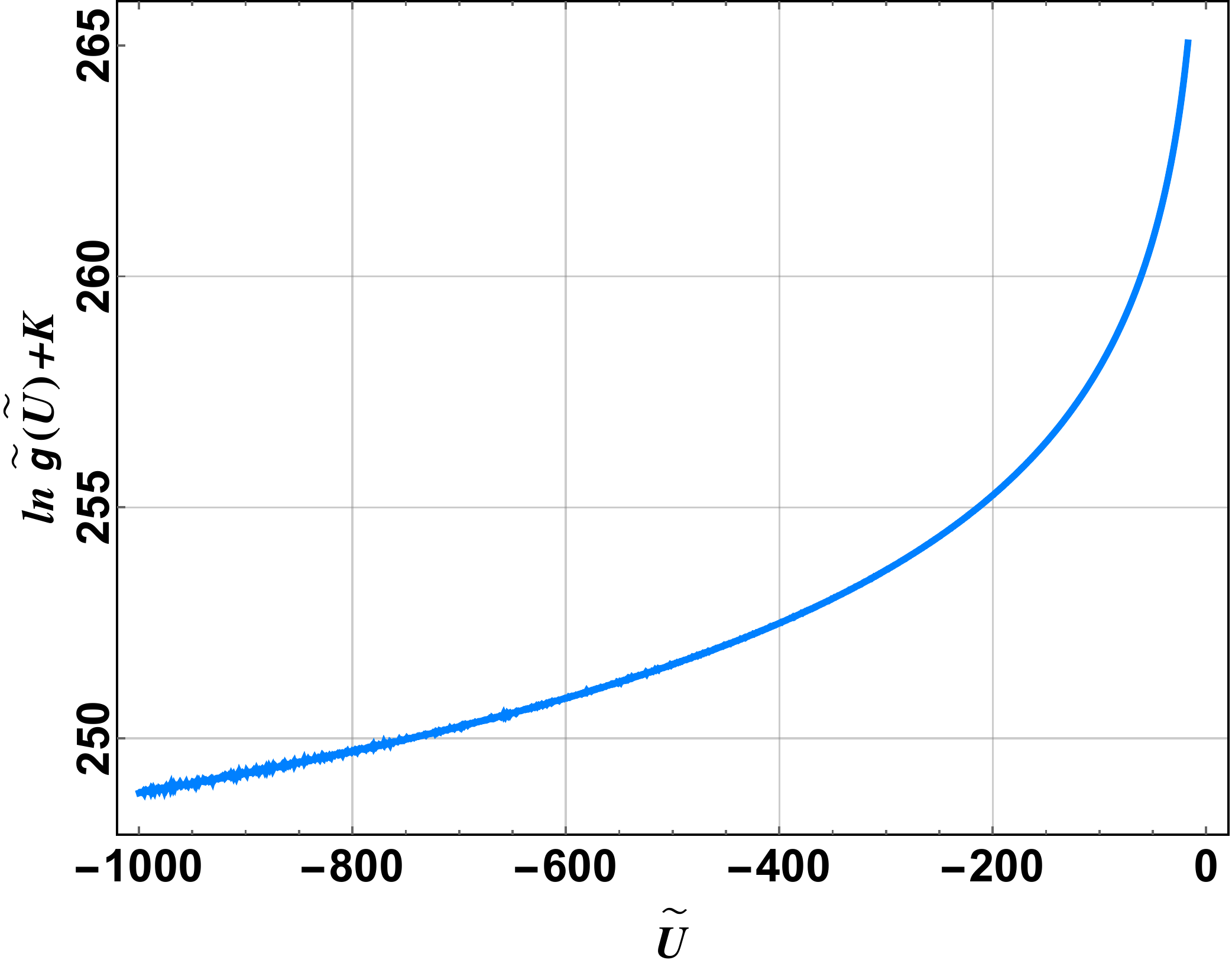}
\caption{Plot of the logarithm of density of states for the potential energy $\Ut$ corresponding to Eq.~\eqref{UBS} for the binary star model.}
\label{Bstar}
\end{figure}

In Fig.~\ref{Bstar}, we show the logarithm of $\gt(\Ut)$ up to an additive constant. The actual value of $\ln \gt(\Ut)$ obtained in the simulation varied from 19\,249 to 19\,275 from which we subtracted 19\,000 from each of the bin values. The resulting data have been plotted in this figure. 
\begin{figure}[!h]
\centering
\includegraphics[width=0.7\linewidth]{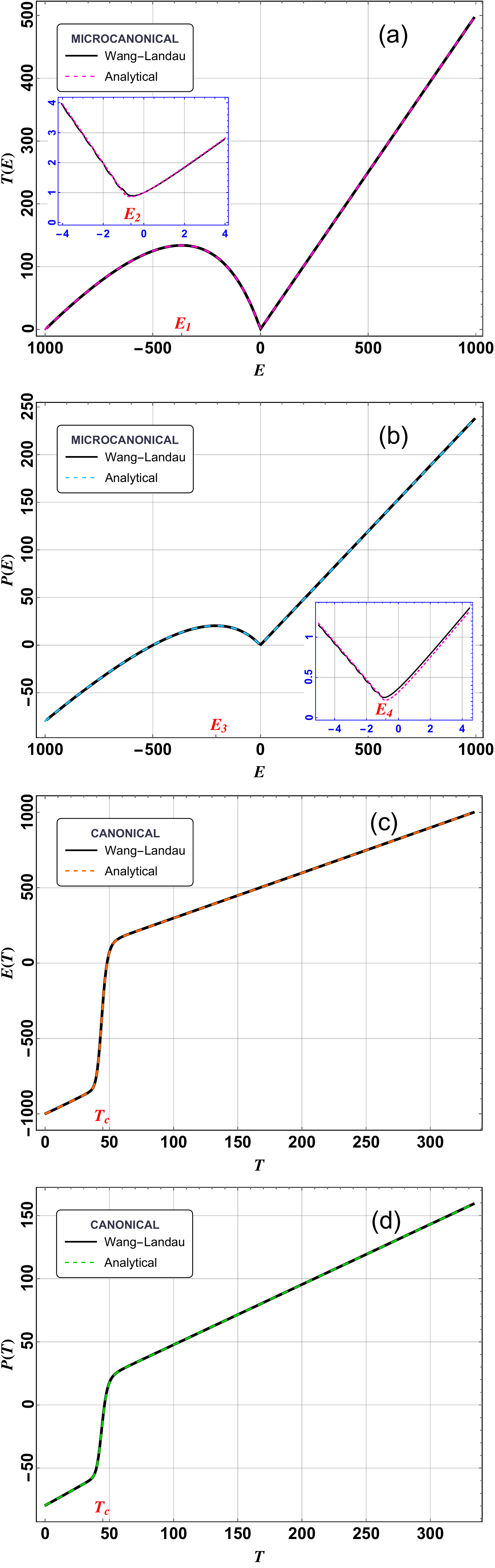}
\caption{Comparison between the Wang-Landau simulation and the analytical results for the binary star model. Subfigures (a) and (b) show temperature and pressure as functions of energy in the microcanonical approach, and (c) and (d) illustrate the average energy and pressure versus the temperature in canonical description.}
\label{BstarObs}
\end{figure}
With the numerical result for $\gt(\Ut)$ available, we can use Eqs.~\eqref{phisum} and~\eqref{psisum} to evaluate the $\phi(\Et)$ and $\psi(\Tt)$ functions, and eventually the thermodynamic observables.
In Fig.~\ref{BstarObs} we show comparison between the analytical results obtained using Eqs.~\eqref{BstarW} and~\eqref{BstarZ} and the results based on Wang-Landau simulation. The subfigures include the temperature, pressure as functions of energy in the microcanonical formalism, and average energy, pressure as functions of temperature in the canonical approach. The specific heat at constant volume can be obtained using the relation $C_v=(\partial T/\partial E)^{-1}_V$ for the microcanonical ensemble and $C_v=(\partial E/\partial T)_V$ for the canonical ensemble. A comparison between the analytical results and the Wang-Landau simulation results are shown in Fig.~\ref{Cv}. We find excellent agreement in all these plots. The small deviations observed in the insets of Figs.~\ref{BstarObs} and ~\ref{Cv} can be attributed to the discretization of the potential energy window and the convergence threshold used in the Wang-Landau simulation.  

\begin{figure}[!t]
\centering
\includegraphics[width=0.7\linewidth]{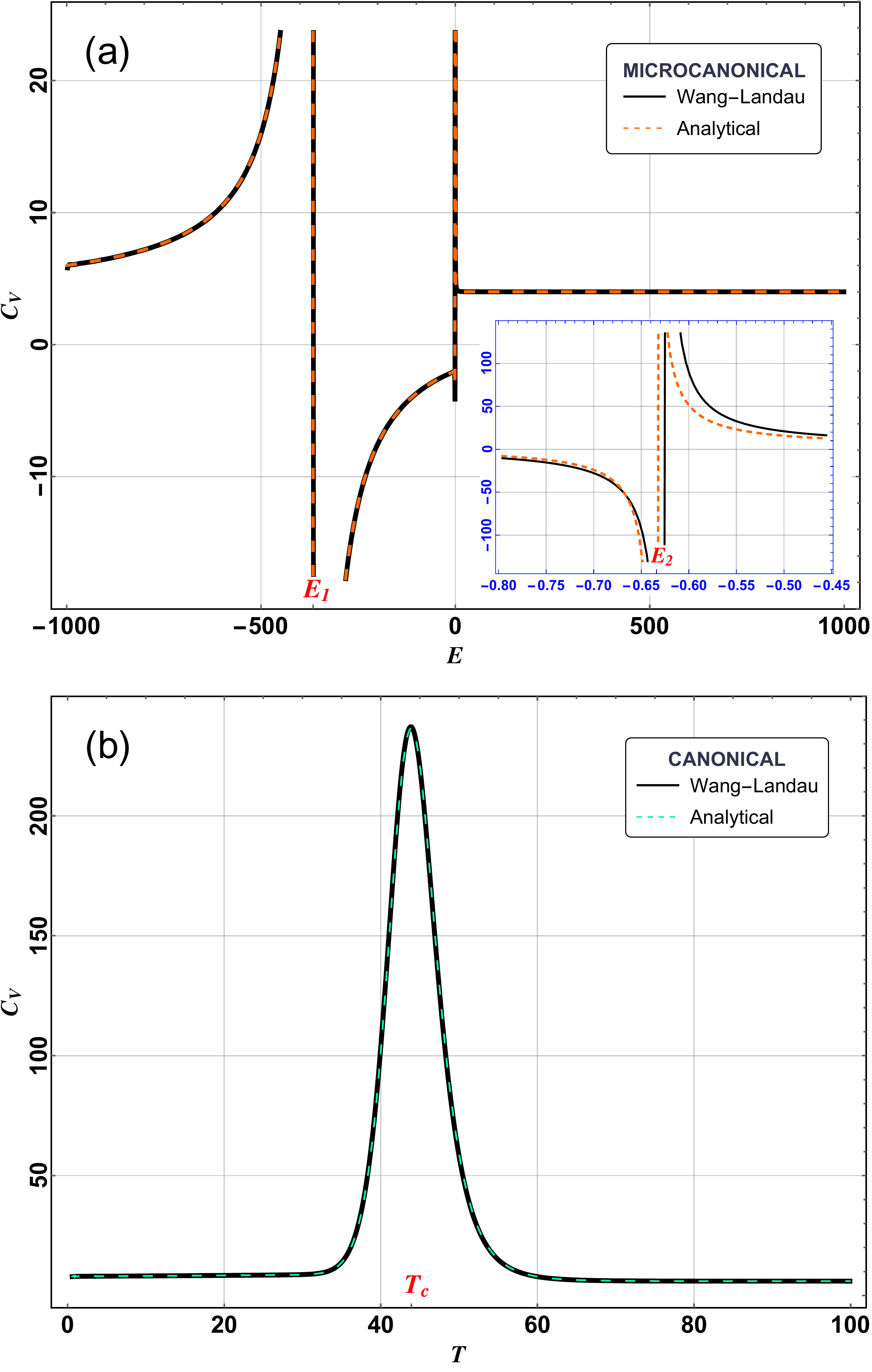}
\caption{Behavior of the specific heat at constant volume ($C_V$) for the binary star system in (a) microcanonical description, and (b) canonical description: Comparison between the Wang-Landau simulation and the analytical results.}
\label{Cv}
\end{figure}

In the plot of temperature versus energy in the microcanonical ensemble, Fig.~\ref{BstarObs}(a), for the lowest energy possible for the system, $E_0=-\Ec/\epsilon=-1000$, the temperature $T$ vanishes. We have introduced here $\Ec=Gm^2/L$. For $E$ close to this value, the system is in a low temperature tightly bound phase with negligible random thermal motion~\cite{Padmanabhan1990}. By examining the local extrema of the temperature function using Eq.~\eqref{Tmc} we observe the following. With increasing energy, the temperature increases until $E$ hits $E_1=-(\sqrt{3}-1)\Ec/(2\epsilon)\approx-366$ after which it starts to decrease until $E_2=-[(3+\sqrt{3})\epsilon+(3-\sqrt{3})]\Ec/[2(\epsilon^2+\epsilon+1)]\approx-0.636$; see the inset. Therefore, the system exhibits positive specific heat in the region $E_0<E<E_1$, and negative specific heat in the region $E_1<E<E_2$, as also seen in Fig.~\ref{Cv}(a). For $E=E_2$ onward, the temperature again starts to increase with an increase in energy, thus indicating another phase with positive specific heat. This is the high temperature phase where thermal fluctuations dominate~\cite{Padmanabhan1990}. From Eq.~\eqref{Pmc} we also see that the system exhibits negative pressure for $\zeta<-1/(2\epsilon)$, i.e., $E<-\Ec/(2\epsilon)=E_0/2=-500$. This can be seen in Fig. 2 (b). Consequently, the system sucks on the walls and wants to contract. Moreover, as $E$ crosses $E_1$ towards higher values, unlike the temperature, the pressure keeps on increasing until $E_3=-(\sqrt{2}-1)\Ec/(2\epsilon)\approx-207$. After this point, the pressure decreases until $E$ reaches $E_4=-[(3-\sqrt{2})+(3+\sqrt{2})\epsilon]\Ec/[2(\epsilon^2+\epsilon+1)\approx-0.794$. Eventually, it increases indefinitely as $E$ is increased beyond this point. From the analytical expression, we find that, if $\epsilon\to0$, then $E_0,E_1\to -\infty$, $E_2\to-(3-\sqrt{3})\Ec/2\approx -0.634$ and therefore, as noted in Ref.~\cite{Padmanabhan1990}, in this limit there is no low temperature region with positive specific heat. Moreover, there is no region of negative pressure for $\epsilon\to 0$ since $E_0,E_3\to-\infty$. Additionally,  $E_4$ moves to $-(3-\sqrt{2})\Ec/2\approx-0.793$.

In the canonical description, in the plot of average energy versus temperature, i.e., Fig.~\ref{BstarObs}(c), as temperature is increased from 0, the average energy increases from the ground state value $-\Ec/\epsilon$. A phase transition occurs at temperature $T_c$, and the energy of the system increases rapidly as it crosses this point. Eventually, the system is pushed into the high temperature phase. Moreover, the pressure changes from negative to positive while crossing the critical point as observed in Fig.~\ref{BstarObs}(d).  In contrast to the microcanonical ensemble, the specific heat is never negative in the canonical ensemble~~\cite{Padmanabhan1990}. Instead we witness a phase transition depicted by a peak in the specific heat curve, as evident from Fig.~\ref{Cv} (b). From the analytical result, by locating the maximum of $C_v$, we find that $T_c\approx 43.886$. This is close to the value of $-\Ec/(3\epsilon \ln\epsilon)\approx 48.255$ which is predicted by an approximate analysis of the partition function in Ref.~\cite{Padmanabhan1990}.

From the above discussion, it is clear that peculiarities associated with the long-range interaction show up in such a simple system and sharply contrasting predictions are made from the microcanonical and the canonical ensemble descriptions~\cite{Padmanabhan1990,Chavanis2002}. An overall plot with $E$, $T$, and $P$ together has been shown in Fig. 4 for both ensembles. Although we have considered here $N=2$ and $V=4\pi/3$, the disagreements between the two statistical ensembles persist even in the thermodynamic limit of $N,V\rightarrow \infty$, with $N/V^{1/3}$ kept fixed as shown in Refs.~\cite{VS2000,VS2002}. 
 \begin{figure}[!t]
\centering
\includegraphics[width=0.7\linewidth]{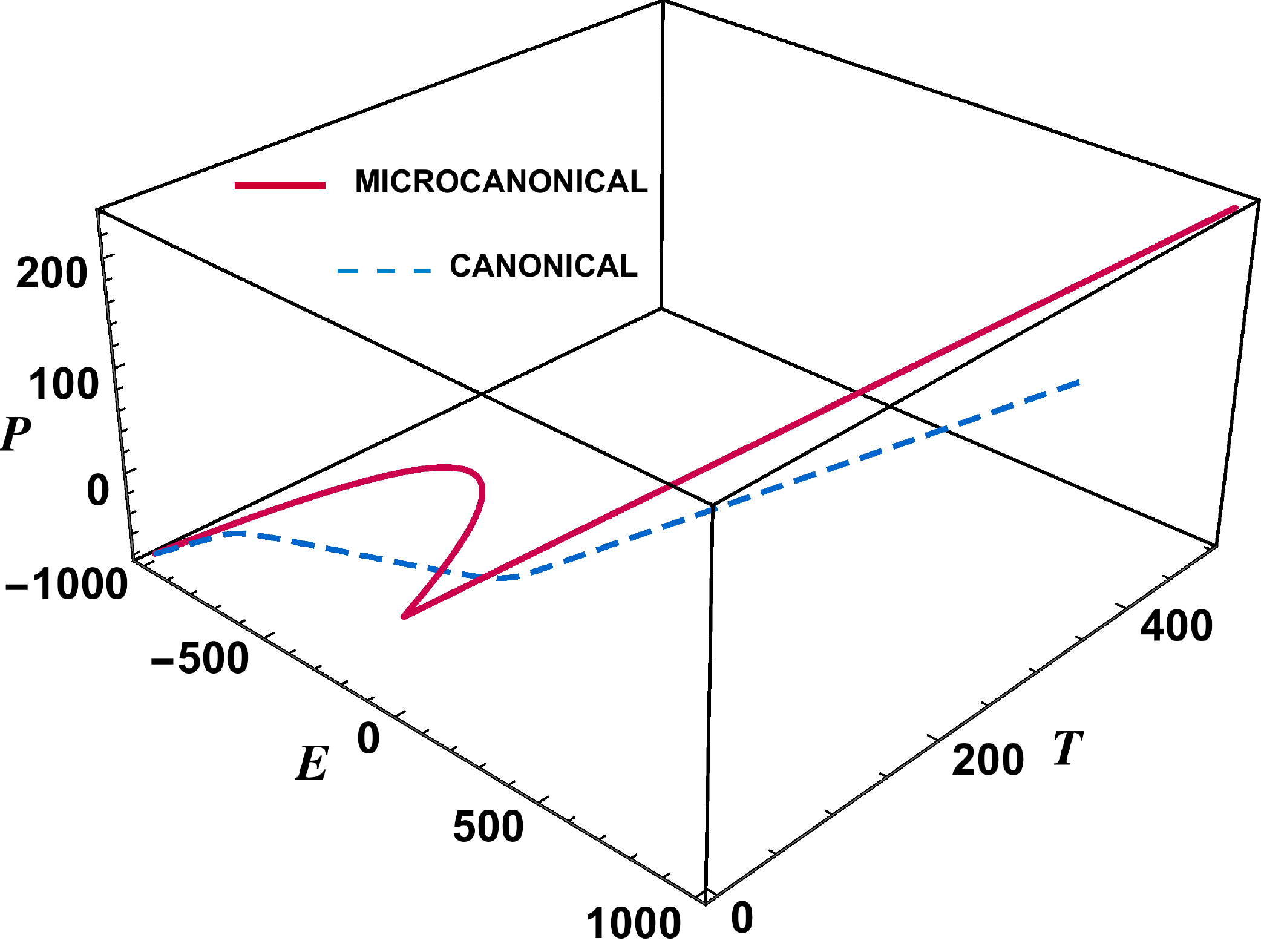}
\caption{Combined plot of energy ($E$), temperature ($T$), and pressure ($P$) for the binary star system in both microcanonical and canonical descriptions.}
\end{figure}

\subsection{Gas in a harmonic trap}

In this model we consider an ideal gas comprising $N$ identical noninteracting particles, each of mass $m$, confined within a cubical box of edge length $L$ and subjected to a harmonic potential. The potential energy function defined inside the box is 
\begin{equation}
\label{Hosc}
\mathcal{U}(\bq_1,...,\bq_N)=\frac{1}{2}m\sum_{j}\omega^2 |\bq_j|^2,
 \end{equation}
 where $\omega$ is the angular frequency. We observe here that $\alpha(L)=L^2$. We position the box such that $\bq_j=\boldsymbol{0}$ coincides with one of the corners of the cubical box placed in the first octant. We may generalize Eq.~\eqref{Hosc} by making $\omega$ dependent on $j$. 
 
We should emphasize that this system is distinct from that of noninteracting particles trapped solely under the harmonic trap and not enclosed in a container. In that particular case, due to the absence of confining walls, the volume and pressure associated with the particles are not well defined, and one has to introduce more appropriate variables to express the thermodynamic relations~\cite{Ligare2010}.
 
 \begin{figure}[!t]
\centering
\includegraphics[width=0.7\linewidth]{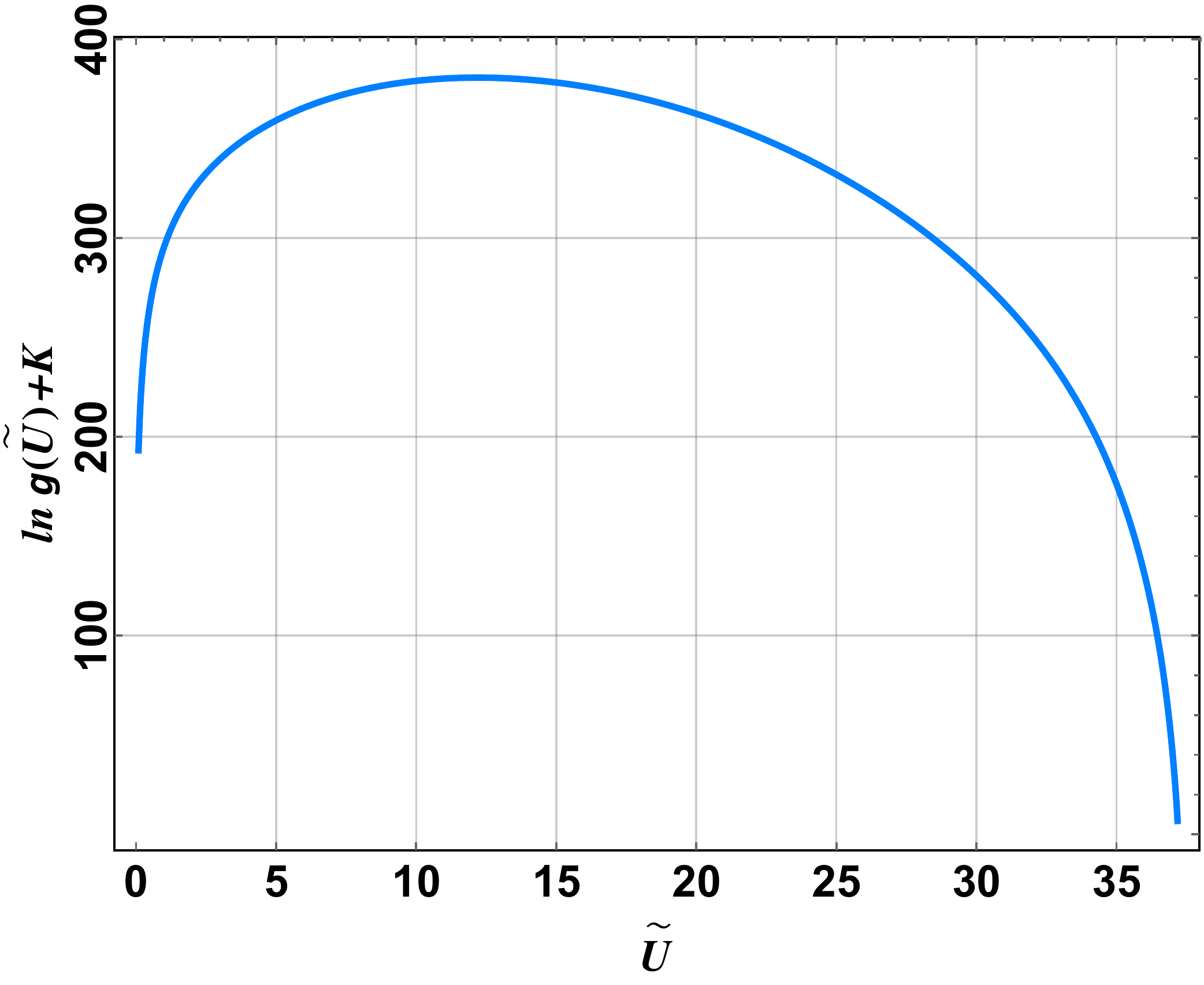}
\caption{Plot of logarithm of density of states for the potential energy $\Ut$ corresponding to Eq.~\eqref{Hosc} for the ideal gas in a harmonic trap within a box.}
\label{hosc}
\end{figure}

For our system, in the microcanonical description, the inverse temperature and the pressure are obtained using Eqs.~\eqref{Tinv} and~\eqref{PbyT}, respectively, as follows:
 \begin{equation}
 \label{HoscTinv}
 \frac{1}{T}=\frac{k_B}{L^2} \phi\left(\frac{E}{L^2}\right),
 \end{equation}
 \begin{equation}
 \label{HoscPbT}
 \frac{P}{T}=\frac{2k_B}{3L^3}\left[3N-1-\frac{E}{L^2}\phi\left(\frac{E}{L^2}\right)\right].
 \end{equation}
 Eliminating $\phi$ from the above two equations yields the equation of state as
 \begin{equation}
 \label{MCanHes}
 PV=\frac{2}{3}(3N-1)k_B T-\frac{2E}{3},
 \end{equation}
 with $V=L^3$.
 On the other hand, in the canonical approach, Eqs.~\eqref{Pt} and~\eqref{Et} yield the pressure and the average energy as follows:
  \begin{equation}
  \label{HoscP}
 P=\frac{2k_BT}{3L^3}\left[3N-\frac{T}{L^2}\psi\left(\frac{T}{L^2}\right)\right],
 \end{equation}
 \begin{equation}
 \label{HoscE}
 E=k_B \frac{T^2}{L^2}\,\psi\left(\frac{T}{L^2}\right).
 \end{equation}
 The equation of state is therefore given by
 \begin{equation}
 \label{CanHes}
 PV=2Nk_B T-\frac{2E}{3}.
 \end{equation}
 We note that, for large $N$, Eqs.~\eqref{MCanHes} and~\eqref{CanHes} are essentially the same.
 \begin{figure}[!t]
\centering
\includegraphics[width=0.7\linewidth]{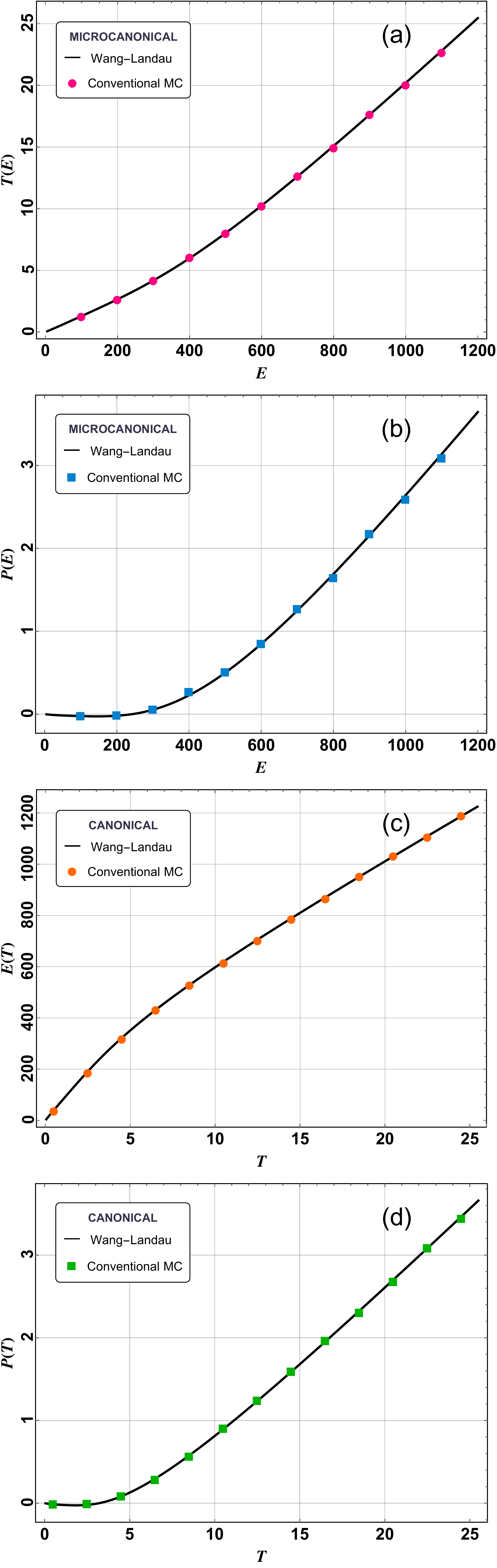}
\caption{Comparison between results based on Wang-Landau and conventional Monte Carlo simulations for the ideal gas in a harmonic trap within a box. Subfigures (a) and (b) show temperature and pressure as functions of energy in the microcanonical approach, whereas (c) and (d) depict the average energy and pressure versus the temperature in the canonical description.}
\label{CanObs}
\end{figure}

 In the present system, since the particles are noninteracting and there is no lower cut-off distance, the lowest possible potential energy is 0 when all the particles accumulate at the origin, i.e., $\bq_j=\boldsymbol{0}$, which coincides with one of the corners of the cubical box. Similarly, the highest potential energy of $3Nm\omega^2 L^2/2$ occurs when all the particles accumulate at the diagonally opposite corner $\bq_j=L(\hat{\bf x}+\hat{\bf y}+\hat{\bf z})$. We examine this system for $N=25$ particles and box of edge length $L=5$. Moreover, we set $m$ and $\omega$ equal to 1.

For the Wang-Landau simulation, we scale the edge lengths and perform the simulation inside a cubical box of edge length 1. Inside this scaled box, the energy window to be explored is $[0,3Nm\omega^2/2]=[0,37.5]$. Since the number of {\it microstates} with potential energies close to the extremal values of 0 and 37.5 is very small, it becomes difficult to explore the energy values close to the extremes if a small bin size is used. For the simulation, we started with a random configuration, and subsequently generated new configurations by perturbing a randomly selected particle by a random amount between $-0.05$ and $0.05$. The time taken for exploring the window $[0.5,37.0]$ with 500 bins was about 30 min, whereas for the window $[0.07,37.2]$ with 1000 bins it took about 60 min. In both cases, a final modification factor of $\exp(10^{-7})$ was considered. We found that the former window with 500 bins was sufficient to produce satisfactory results for the thermodynamic observables. However, we present the results below for the latter. The logarithm of density of states is shown in Fig.~\ref{hosc}. The actual values obtained in the simulation varied around $3\times10^6$ from which we have subtracted a common value to obtain this plot. We note a sharp drop in the $\ln \gt(\Ut)$ curve towards the extremes, especially towards the lower end, thereby indicating decreasing number of the associated potential energy {\it microstates}.
 \begin{figure}[!t]
\centering
\includegraphics[width=0.75\linewidth]{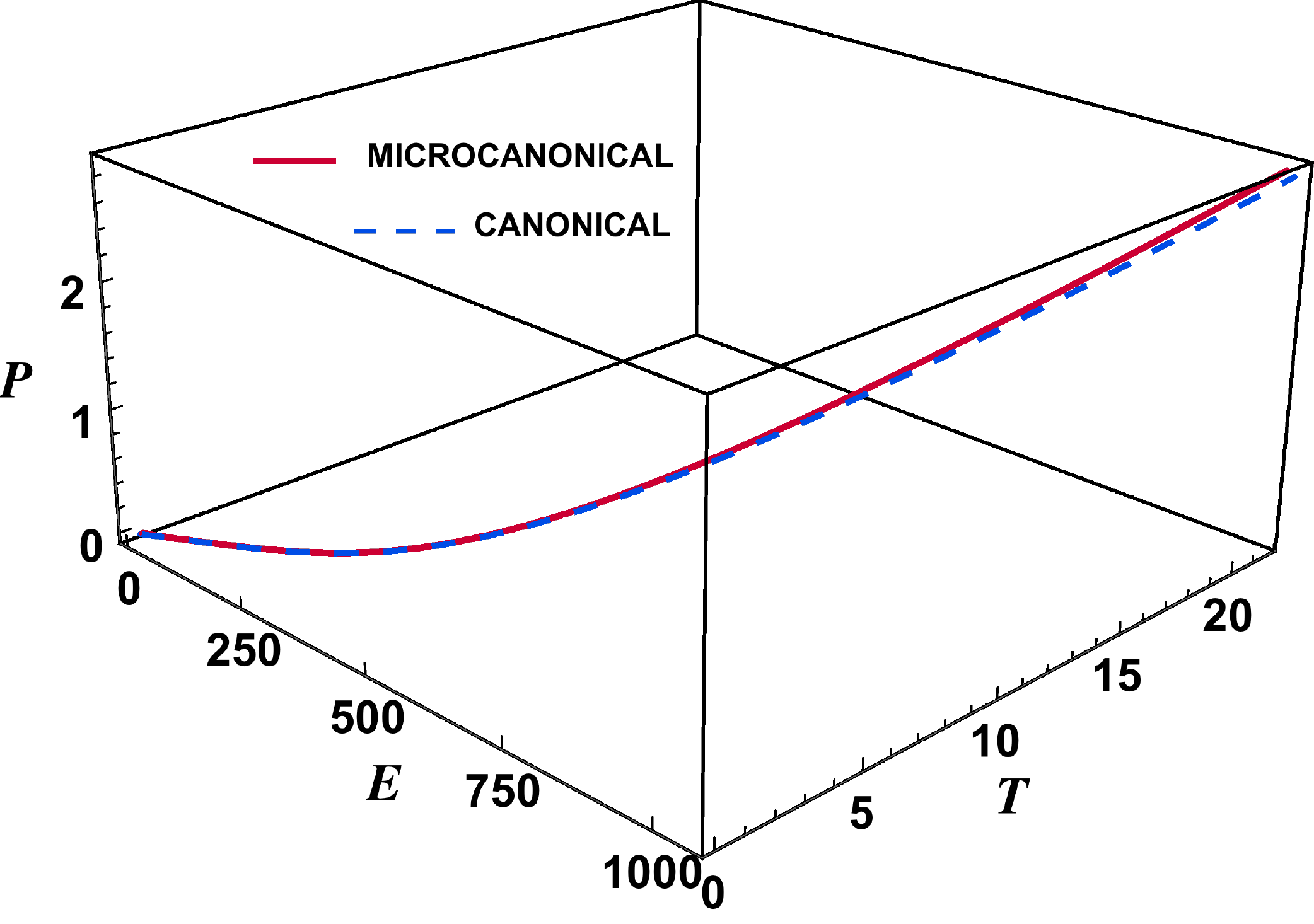}
\caption{Combined plot of energy ($E$), temperature ($T$), and pressure ($P$) for the the ideal gas in a harmonic trap within a box.}
\label{hoscETP}
\end{figure}

The observables laid out in Eqs.~\eqref{HoscTinv},~\eqref{HoscPbT},~\eqref{HoscP}, and~\eqref{HoscE} can be obtained using $\gt(\Ut)$. For this system, the partition function for the canonical  description can be worked out analytically as $Z(T,V)\propto [T\,\text{erf}(\sqrt{m\omega^2L^2/2kT})]^{3N}$, where erf($\cdot$) represents the error function~\cite{GR2007}. Consequently, the associated thermodynamic observables can be calculated. However, it does not seem feasible to obtain a closed form analytical result for the entropy $S(E,V)$ within the microcanonical description. Therefore, we rely on conventional Monte Carlo simulations to validate the Wang-Landau results. For the canonical ensemble, we use the Metropolis algorithm with Boltzmann-Gibbs factor $\exp[-\mathcal{U}/(k_B T)]$ as the statistical weight and perform the simulation in a box of size $L=5$; see Eq.~\eqref{CanSW}. For the microcanonical ensemble, Eq.~\eqref{MCanSW} implies the weight $(E-\mathcal{U})^{3N/2-1}\Theta(E-\mathcal{U})$ to be implemented in the Metropolis algorithm. Clearly, unlike the Wang-Landau algorithm, we have to perform the simulations individually for each desired energy and temperature values. For the microcanonical ensemble, it took about 65 min for simulation involving 11 energy values. In the case of the canonical ensemble, the time taken was about 30 min for a simulation comprising 13 temperature values. In Fig.~\ref{CanObs}, we see the comparison between the Wang-Landau results and the conventional Monte Carlo schemes as described above. An excellent agreement is found in all cases. We note that the pressure briefly becomes negative in both microcanonical and canonical descriptions and therefore the system tries to contract sucking on the walls of the box. This region corresponds to the potential energy dominating over the kinetic energy. 
 Finally, Fig.~\ref{hoscETP} shows the Wang-Landau-simulation plots of $E$, $T$, and $P$ together for both microcanonical and canonical ensembles. In this case, since there are no peculiarities involved, the two descriptions agree quite well.

\section{Conclusion}
\label{SecSum}

In this paper, we described an efficient way of applying the Wang-Landau algorithm to systems which satisfy certain length (or equivalently volume) dependent scaling behavior for their potential energy function defined within the enclosing box. This scaling behavior is rather general and is observed in several important statistical models. Our method allows one to calculate thermodynamic observables that involve energy, temperature, and volume fluctuations, appropriate to the microcanonical or the canonical description. Interestingly, this requires obtaining the density of states corresponding to the potential energy part only by performing the Wang-Landau simulation in a one-dimensional space for a scaled box of unit size. Our approach is quite advantageous compared to the conventional Monte Carlo schemes where one needs to perform simulation for individual energy and temperature values or Wang-Landau scheme aimed to obtain the density of states as a function of both energy and volume. We applied our methodology to study Padmanabhan's binary star model and a noninteracting gas trapped in a harmonic potential within a container and found excellent results in both cases. It would be of interest to investigate how this approach performs in more complicated systems where one lacks any analytical solution.


\begin{acknowledgments}
This work initiated from a project that was carried out at Shiv Nadar University under the Opportunities for Undergraduate Research (OUR) scheme. The authors are grateful to the anonymous reviewers for their fruitful suggestions.
\end{acknowledgments}

\end{document}